\documentclass[aps,prl,twocolumn,amsmath,amssymb]{revtex4} 
 
\usepackage{graphicx}% Include figure files 
\usepackage{dcolumn}% Align table columns on decimal point 
\usepackage{bm}% bold math 
\def\lsim{\raise0.3ex\hbox{$<$\kern-0.75em\raise-1.1ex\hbox{$\sim$}}} 
\def\gsim{\raise0.3ex\hbox{$>$\kern-0.75em\raise-1.1ex\hbox{$\sim$}}}

\newcommand{\beq}{\begin{equation}} 
\newcommand{\eeq}{\end{equation}} 
\newcommand{\bea}{\begin{eqnarray}} 
\newcommand{\eea}{\end{eqnarray}}

\begin{document} 
 
\title{Dynamical freeze-out condition in ultrarelativistic heavy ion collisions} 
 
\author{ 
K.J. Eskola, H. Niemi and P.V. Ruuskanen} 
\email{kari.eskola, harri.niemi, vesa.ruuskanen@phys.jyu.fi} 
\affiliation{Department of Physics, 
P.O.Box 35, FIN-40014 University of Jyv\"askyl\"a, Finland\\ 
Helsinki Institute of Physics, 
P.O.Box 64, FIN-00014 University of Helsinki, Finland}

\date{24.10.2007}

\begin{abstract} 
We determine the decoupling surfaces for the hydrodynamic description of
heavy ion collisions at RHIC and LHC by comparing the local hydrodynamic
expansion rate with the microscopic pion-pion scattering rate. 
The pion $p_T$ spectra for nuclear collisions at RHIC and LHC are computed by applying the
Cooper-Frye procedure on the dynamical-decoupling surfaces, and compared
with those obtained from the constant-temperature freeze-out surfaces.
Comparison with RHIC data shows that the system indeed decouples when the
expansion rate becomes comparable with the pion scattering rate. The
dynamical decoupling based on the rates comparison also suggests that the
effective decoupling temperature in central heavy ion collisions remains
practically unchanged from RHIC to LHC.

\end{abstract} 
 
%\pacs{25.75.-q, 12.38.Mh, 47.75.+f, 24.85.+p} 
 
\maketitle 

\section{Introduction}
Non-dissipative hydrodynamical description has been quite successful in explaining the data 
on low-$p_T$ transverse momentum spectra and elliptic flow in nucleus-nucleus collisions at RHIC 
\cite{Hirano:2001yi,Hirano:2001eu,Hirano:2002ds, Eskola:2002wx,Kolb:2003dz,Huovinen:2003fa,Eskola:2005ue,Huovinen:2006jp}. 
First results from viscous hydrodynamics suggest, that the viscosity~\cite{Baier:2006gy, Romatschke:2007jx, Romatschke:2007mq, Song:2007fn} 
of quark-gluon plasma should be small to accommodate the data, in particular those on elliptic flow. Apart from the viscous effects,
the main uncertainties in the hydrodynamic modelling of heavy ion collisions lie in the computation 
of the initial conditions, the choice of equation of state (EoS) for the QCD matter and, 
the subject of this paper, the treatment of the freeze-out from a local thermal equilibrium state to 
free hadrons. The standard way of handling the freeze-out is the Cooper-Frye decoupling 
procedure~\cite{CooperandFry}. Also hadron cascade models have been successfully implemented to 
describe late hadronic dynamics~\cite{Teaney:2001av, Hirano:2005xf}, but typically the initial hadron 
distributions in these simulations are also calculated using the Cooper-Frye method.

Traditionally, the transverse momentum distributions of hadrons at freeze-out are calculated as 
emissions from a constant-temperature hypersurface, with the value of the decoupling temperature 
$T_{\rm dec}$ fixed from data. Since the constant-$T_{\rm dec}$ Cooper-Frye decoupling 
procedure is not based on scattering rates, it cannot predict the freeze-out temperature or conditions for
the LHC Pb+Pb collisions. To obtain predictions for hadron $p_T$ spectra at the LHC, one often 
assumes -- like we have done in \cite{Eskola:2005ue} -- that $T_{\rm dec}$ at RHIC and LHC are equal.

In this work, we shall study this uncertainty by determining the decoupling surface by using 
a dynamical freeze-out condition instead of a constant $T_{\rm dec}$. That is, we compare the local 
hydrodynamical expansion rate with the local scattering rate of pions, and define the Cooper-Frye 
freeze-out surface to be where the expansion rate exceeds the scattering rate, possibly multiplied by 
a constant. To our knowledge, this idea was first discussed in Refs.~\cite{Dumitru:1999ud, Hung:1997du}.
Impact parameter dependence of freeze-out conditions using the same idea is studied in Ref.~\cite{Heinz:2007in}.
The ratio of these two rates can be fixed on the basis of the measured pion $p_T$ spectra at RHIC. 
Because this ratio now depends only on the local properties of the system (temperature and flow) 
it can well be expected to remain unchanged in the extrapolation to higher collision energies.
We are thus able to predict the freeze-out surface for the bulk of QCD matter in Pb+Pb collisions at the LHC 
based on the dynamics of the system.

We calculate pion $p_T$ spectra using freeze-out surfaces from the dynamical decoupling condition, 
and compare the results with the constant-$T$ decoupling at RHIC and LHC. 
Interestingly, two main results come out of this exercise: 
First, comparison with the pion $p_T$ spectra at RHIC nicely confirms the expectation that the 
expansion and scattering rates are of the same order at freeze-out.
Second, the local dynamical freeze-out condition leads to practically the same effective decoupling temperature 
at LHC as at RHIC which is non-trivial since the evolution of flow is quite different at the two collision 
energies. 

\section{Hydrodynamic setup and rates}
A detailed account of the hydrodynamic setup we use can be found in \cite{Eskola:2005ue}. 
We consider boost-invariant perfect fluid hydrodynamics with transverse expansion for azimuthally  
symmetric systems, i.e. study (nearly) central collisions only. The EoS is given by the bag model, 
where the quark-gluon plasma (QGP) is an ideal gas of massless quarks, antiquarks and gluons, and the 
hadron  resonance gas (HRG) ~\cite{Sollfrank_prc} consists of all hadronic states up to a mass 2 
GeV~\cite{PDG}. Critical temperature is fixed to $T_c = 165$ MeV at $\mu_B = 0$. 
Full thermalization, both kinetic and chemical, until decoupling is assumed, after which we also 
include the 2- and 3-body resonance decays of unstable hadronic states. 

The initial conditions for the hydrodynamic evolution -- the total energy and net-baryon number deposited 
into a rapidity unit by an initial time -- for RHIC and LHC $A+A$ 
collisions, are obtained from the pQCD + (final state) saturation model ~\cite{EKRT} as explained in 
detail in \cite{Eskola:2005ue}. In our previous studies~\cite{Eskola:2005ue, Eskola:2002wx} the
initial energy density was taken to be proportional to the density of nucleon binary collisions (BC). 
To probe the uncertainty due to the initial transverse geometry, we study also an initial state where 
the energy density scales with the density of wounded nucleons (WN). In the latter case, the initial 
entropy is kept the same as in the BC case.

A measure of the hydrodynamical expansion rate is given by the expansion scalar
\begin{equation} 
\theta = \partial_\mu u^\mu = \frac{\partial \gamma_{r}}{\partial \tau} + \frac{\gamma_r}{\tau} + \frac{\partial (\gamma_r v_r)}{\partial r} + \frac{\gamma_r v_r}{r},
\end{equation}
where $u^\mu = \gamma(1, v_r, v_\phi, v_z) = \gamma_r(\cosh \eta, v_r, 0, \sinh \eta)$ is the flow 
four-velocity of the matter, $\eta$ is the coordinate space rapidity and $\tau$ the longitudinal proper time.
This can be interpreted as the rate of relative volume change $\dot{V}/V$ in the local rest frame \cite{Dumitru:1999ud},  
with $\dot{V}= \partial V/\partial \tau'$, where $\tau'$ is the fluid element's proper time.
An easy way to see this is to express entropy conservation in the local rest frame as
\begin{equation} 
\dot{s} = -s\theta,
\end{equation} 
where $s$ is the entropy density. From $s=S/V$ it follows 
\begin{equation} 
\theta = -\frac{\dot{s}}{s} = \frac{\dot{V}}{V},
\end{equation} 
where the last step follows from $S =$ constant.
The expansion rate $\theta$ can be easily calculated once the hydrodynamic solution is known. 

As long as the microscopic collision rate per particle, $\Gamma$, is large relative to the expansion 
rate, $\Gamma\gg\theta$, the system can be assumed to be in local thermal equilibrium. 
Since we are here interested in the bulk matter properties only, and since without studying a 
separate chemical decoupling of heavier particles we cannot reproduce their $p_T$ spectra in detail
\cite{Eskola:2005ue}, we consider only pion-pion collisions in the scattering rate. We thus  
approximate the local collision rate per particle as 
\begin{equation}
\Gamma = n_{\pi}(T)\, \sigma_{\pi\pi}(T),
\end{equation}
where the density of (massless) pions is $n_{\pi}(T) = 3 \frac{\zeta(3)}{\pi^2}T^3$.
For the average pion-pion cross-section at a given temperature $T$, we use the following 
parametrization, 
\begin{equation}
\sigma_{\pi\pi}(T) = \sigma_{0} + \frac{C\,T^{2}}{(T - T_{0})^2 + (\Delta/2)^2},
\end{equation}
where $\sigma_{0} = 0.60$ fm$^2$, $C = 0.78$ fm$^2$, $T_{0} = 105$ MeV and
$\Delta = 170$ MeV, which reproduces the results of Ref.~\cite{Bertsch:1987ux} in the temperature 
range $T<165$~MeV of the HRG phase.
Note in particular, how due to the rho resonance peak the cross section 
decreases with decreasing $T$, instead of a naive dimensional $T$-scaling $\sigma_{\pi\pi}\sim T^{-2}$. 
Note also that we do not need to consider partonic scattering rates at all. Since we are interested 
in the decoupling of final hadrons, we can limit the rate study to the HRG phase only.
In the Bjorken picture \cite{BJORKEN} of heavy ion collisions, which is adopted here, 
the hydrodynamic expansion starts with very strong
longitudinal expansion. In the boost-invariant case without transverse expansion, 
the fluid velocity, $u^\mu=(\cosh\eta,0,0,\sinh\eta)$, 
depends only on the coordinate space rapidity $\eta$,
and the (longitudinal) expansion rate is 
given by  the inverse of the longitudinal proper time $\tau$,  
$\theta=1/\tau$. During the expansion, pressure gradients develop a 
strong transverse flow, which leads to an increasing transverse expansion rate.
The decreasing longitudinal and the increasing transverse expansion rates lead to a 
system where the local expansion rate is high at the spatial boundary of the 
system, where transverse gradient is strong, while near the center of 
the fireball the expansion rate is relatively weak.
%%%%%%%%%%%%%%%%%%%%% FIGURE %%%%%%%%%%%%%%%%%%%%%%%%%%%%%%%% 
\begin{figure}[bht] 
\vspace{-0.0cm} 
\hspace{-0.0cm} 
\includegraphics[height=8.5cm]{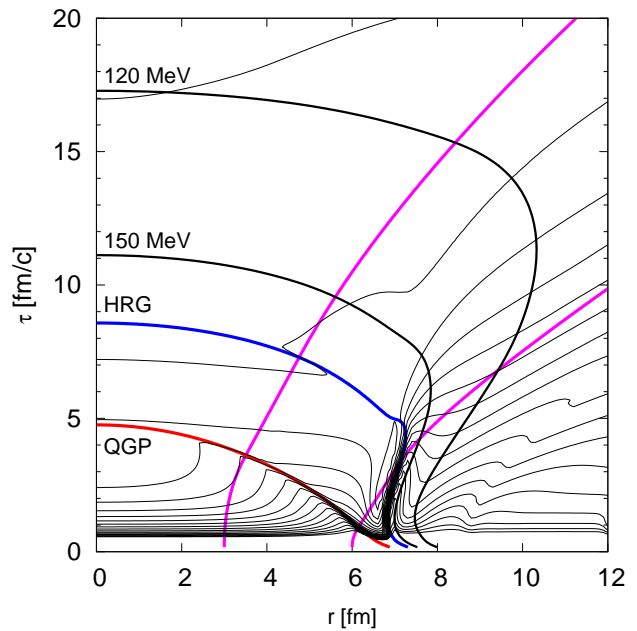} 
\label{thetaconst} 
\vspace{-0.2cm} 
\caption{\protect\small  Contours of local expansion rate $\theta$, 
phase boundaries and two isotherms in 5 \% most central RHIC 200 GeV Au+Au collision
for BC initial state. Reading from the top of the figure, $\theta = 0.2, 0.3, \dots 1.8$.
Also two flowlines, starting from $r=3$ fm and $r=6$ fm, are shown.} 
\end{figure} 
%%%%%%%%%%%%%%%%%%%%% FIGURE %%%%%%%%%%%%%%%%%%%%%%%%%%%%%%%% 
%%%%%%%%%%%%%%%%%%%%% FIGURE %%%%%%%%%%%%%%%%%%%%%%%%%%%%%%%% 
\begin{figure}[ht] 
\vspace{-0.0cm} 
\hspace{-0.0cm} 
\includegraphics[height=6.0cm]{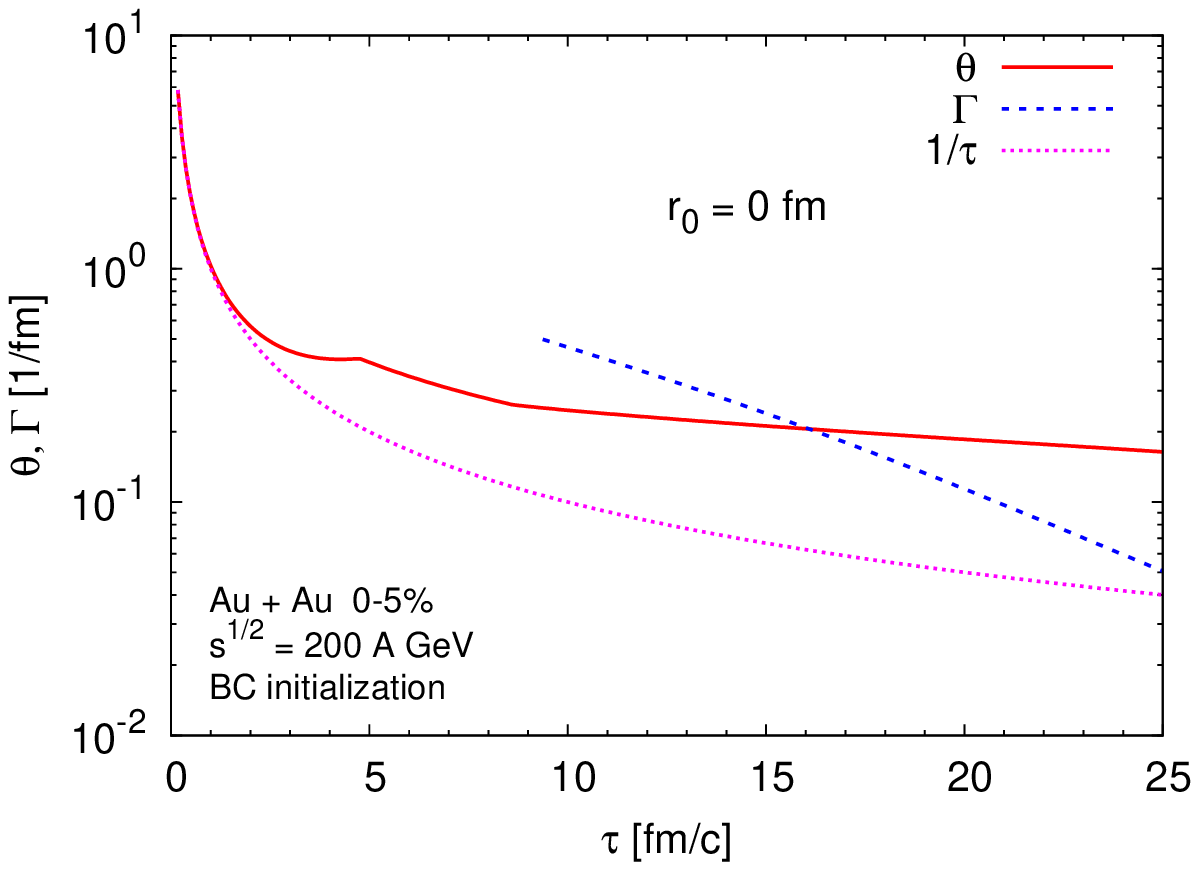} 
\includegraphics[height=6.0cm]{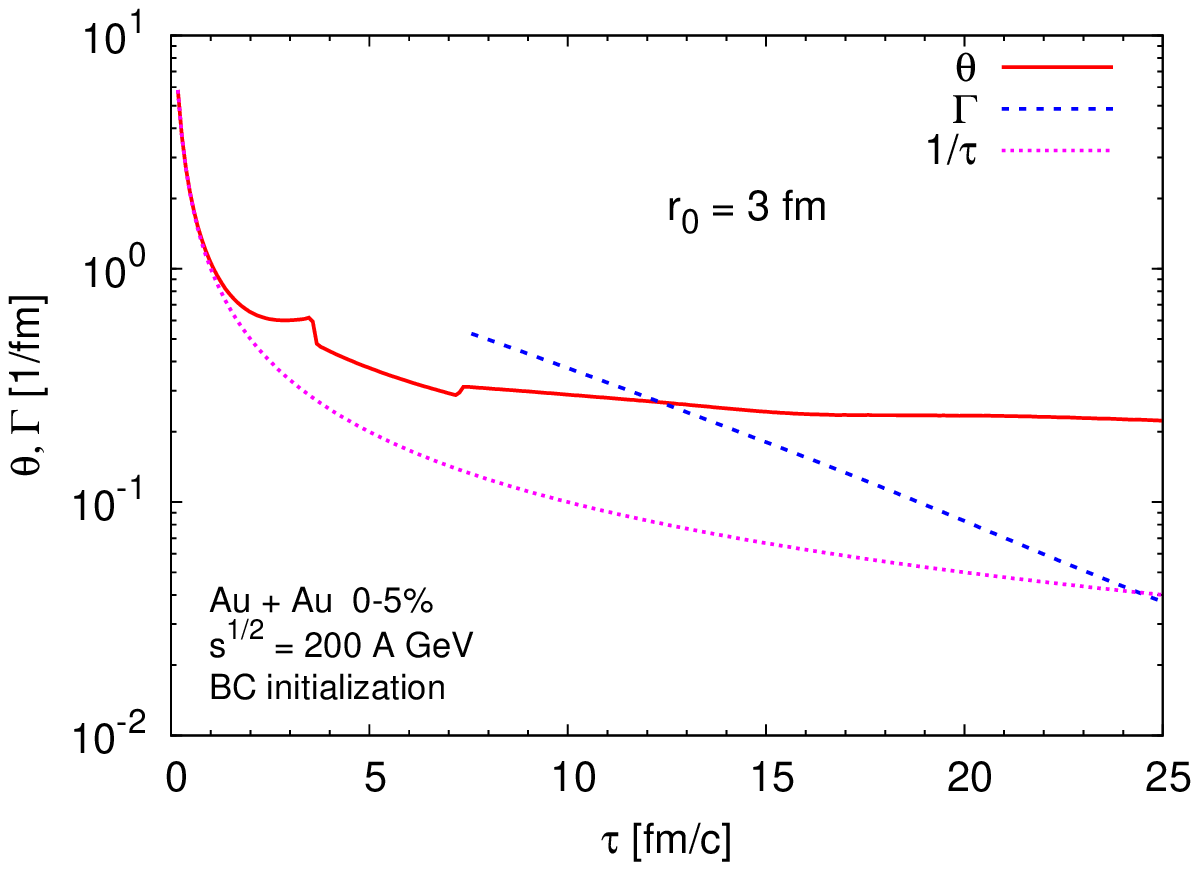} 
\includegraphics[height=6.0cm]{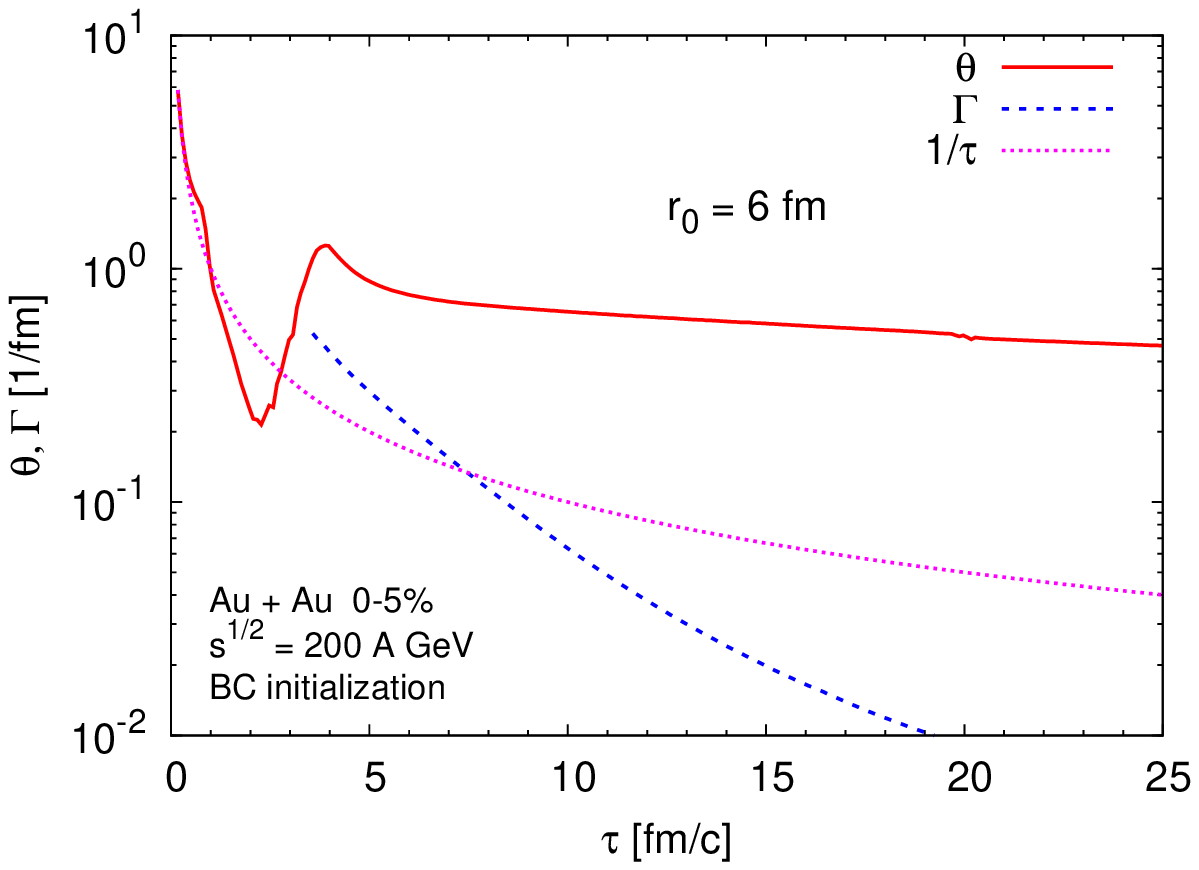} 
\label{theta_and_rate} 
\vspace{-0.0cm} 
\caption{\protect\small Expansion and scattering rates along the flowlines starting from $r = 0,\,3$ 
and $6$ fm for RHIC 200 GeV Au+Au collision. Flowlines are shown in Fig.~1. Also the 
expansion rate $\theta = 1/\tau$ for a system with longitudinal boost invariant expansion only is shown.} 
%\caption{\protect\small} 
\vspace{-0.4cm} 
\end{figure} 
%%%%%%%%%%%%%%%%%%%%% FIGURE %%%%%%%%%%%%%%%%%%%%%%%%%%%%%%%% 

In Fig.~1 we show the contours of constant $\theta$ for a nearly central RHIC 
Au+Au collision at $\sqrt{s_{NN}}=200$ GeV together with the phase boundaries and 
two isotherms $T=150$ MeV and $T=120$ MeV. Of the two isotherms shown, the one with 
$T=150$ MeV corresponds to the freeze-out surface, which leads to an agreement with 
the RHIC data on pion $p_T$ spectra, as discussed in \cite{Eskola:2005ue}.
Also two flowlines starting from transverse locations $r=3$ fm and $r=6$ fm are drawn 
in the figure. 

Figure~2 shows the computed expansion and scattering rates  
at $r = 0$ fm and along the two flowlines in Fig.~1. For comparison with the case of no 
transverse expansion, also the expansion rate $\theta = 1/\tau$, due to the longitudinal boost 
invariant flow only, is shown. As seen in Fig.~2, the system starts its evolution with longitudinal 
but without transverse expansion and $\theta$ drops like $1/\tau$ until the increasing transverse expansion starts to 
contribute to the rate, slowing down its decrease. Interestingly, this can even develop a local 
maximum of $\theta$ near the boundary of QGP and mixed phase. In Fig.~1 the rarefaction wave 
from the system boundary can be seen as a bending of the $\theta$ contours 
towards the center of the fireball in the QGP phase. The same phenomenon can be seen in Fig.~2
when the expansion rate starts to exceed the longitudinal expansion rate.

Because pressure gradients are stronger near the boundary of the system, the transverse expansion
rate grows faster at larger values of transverse radius $r$. This is seen clearly in Fig.~1,
where in the QGP phase the expansion rate grows as a function of $r$ at constant $\tau$.
In the mixed phase pressure gradients vanish, which stops the increase of the expansion rate.
In contrast to the behaviour in the QGP phase, in the mixed phase the expansion rate decreases as a function of $r$
at constant $\tau$. This is because the matter has spent less time in the QGP
phase at larger values of $r$, thus there has been less time for transverse flow 
to develop. When the system goes through the phase transition into HRG, the shock wave
formed generates very strong gradients near the phase boundary in the HRG, and this in 
turn leads again to a strong increase of the transverse expansion as a function of $r$ in the HRG phase.

In summary, the phase transition first slows down the increase of the transverse expansion 
rate, but this is later counteracted by resulting strong gradients in hadron 
gas near the phase boundary. This is seen in Fig.~1, where, as a result
of the phase transition, the $\theta$ contours are compressed near the mixed phase and 
hadron gas boundary. This effect will be important when we construct 
the decoupling boundaries by comparing the local scattering rate to the local expansion 
rate. 

In Fig.~2, for the flowline starting from $r = 6$ fm (see the lowest panel), we see that $\theta$ 
can actually decrease below the longitudinal value $1/\tau$ for a short while during the phase 
transition. This is due to a transverse compression rather than expansion in a small region
near the boundary during the phase transition, caused by the QGP accelerating against
the mixed phase where the pressure gradients are zero and matter is not accelerating.
This is how compressional shock wave is mechanically formed between the QGP and the HRG phases.
Same phenomenon can be seen in Fig.~1, where expansion rate has a strong local minimum in the
mixed phase near the edge of the fireball.

Because the scattering rate is a function of temperature alone, the constant-temperature contours  
represent also the contours of the scattering rate. It is clear from Fig.~1 
that the contours of expansion and reaction rates do not coincide. Both the 
scattering and expansion rates decrease in general with time, but the decrease of the scattering rate is
faster -- thanks to the resonance peak in the pion scattering cross section. 
At sufficiently low temperatures, this eventually leads to the situation 
where the scattering rate drops below the expansion rate, as illustrated 
in the two top panels of Fig.~2, and this is when the system is expected to decouple. 
Thus, the decoupling surface is given by the crossing points of the expansion and 
scattering rate lines like in the two top panels of Fig.~2.

The bottom panel of Fig. 2 shows the situation near the edge of the system, where 
the scattering rate in the HRG falls below the expansion rate already at the beginning of the HRG phase.
In this case, we assume the decoupling to take place at the boundary of the mixed phase and the HRG.

Figure 3 shows the phase boundaries and constant temperature contours as in Fig.~1, 
and in addition also the decoupling boundaries from a dynamical condition 
\begin{equation}
\theta = c \Gamma,
\label{dec_crit}
\end{equation} 
with different values of the constant $c$. (The crossing points in Fig. 2 correspond to $c=1$.)
In the hydrodynamic approach with phase transition we cannot decouple the matter before it has 
completely hadronized. Thus the decoupling curve can be only in HRG or on the boundary of mixed 
phase and HRG. Dynamics of the system with phase transition is such that it always forces decoupling
curves near the mixed phase--HRG phase boundary, as seen in Fig.~1, where contours of expansion rate
are compressed near the boundary. We only need a slight adjustment to force the freeze-out boundaries 
to be strictly outside the mixed phase.
It turns out that for the values of $c$ considered here, the rate of transverse expansion on the 
edge of the matter is so large that the decoupling curve always first follows the boundary of 
the mixed and HRG phases. For the smallest value, $c=0.5$, the decoupling curve (see Fig. 3) 
is close to the mentioned phase boundary at all radial distances, but for larger values of $c$ the curves 
separate when the phase boundary curve turns sharply towards the center of the matter.
Near the center, where the expansion rate is smaller than at the edge since the contribution from 
the transverse expansion is small, the freeze-out time can become large and the freeze-out 
temperature small. Thus along a dynamic decoupling curve the temperature starts from $T_c=165$ MeV 
at the edge of the matter and drops, for $c\gsim 1$ to clearly smaller values of the order 
of $100$ to $120$ MeV as the curve reaches the center. As seen in Fig.~3, the dynamical decoupling curves 
typically cross the constant-temperature curves.
%%%%%%%%%%%%%%%%%%%%% FIGURE %%%%%%%%%%%%%%%%%%%%%%%%%%%%%%%% 
\begin{figure}[bht] 
\vspace{-0.0cm} 
\hspace{-0.0cm} 
\includegraphics[height=8.5cm]{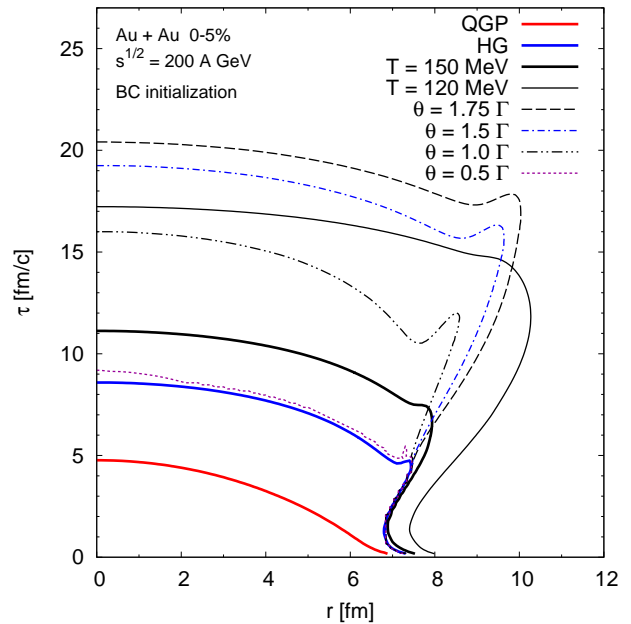} 
\label{flowRHIC_BC} 
\vspace{-0.5cm} 
\caption{\protect\small  Decoupling boundaries from a dynamical condition $\theta = c\Gamma$ 
corresponding to 5 \% most central RHIC 200 GeV Au+Au collisions. Calculation is initialized with BC profile.
Also phase boundaries and constant temperature decoupling boundaries are shown.} 
\vspace{-0.0cm} 
\end{figure} 
%%%%%%%%%%%%%%%%%%%%% FIGURE %%%%%%%%%%%%%%%%%%%%%%%%%%%%%%%% 

For large values of $c$ the rather big differences in temperature along the decoupling curve can cause the following 
problem. Without a chemical freeze-out, at or soon after the formation of hadron phase, 
the density of heavier particles can become much smaller on the flat, space-like part of the decoupling 
curve than on the time-like part at the edge. Since in the Cooper-Frye prescription the contribution to the spectrum from the time-like 
part can become negative at small $p_T$'s, it can happen that too few heavy 
particles are available on the low temperature space-like part to exceed the negative contributions, and the spectrum may remain negative 
close to $p_T=0$. From our point of view there is no reason to go to large values of $c$ since $c\sim 1$ 
is the natural assumption. Also, we are considering only pions in this work, and the described effect is
much weaker for pions.

Finally, we note that so far all the figures shown have been for 
the BC initial state, but we have checked that similar results also hold for the WN profile.

\section{RHIC results}

%%%%%%%%%%%%%%%%%%%%% FIGURE %%%%%%%%%%%%%%%%%%%%%%%%%%%%%%%% 
\begin{figure}[bht] 
%\vspace{0.6cm} 
%\hspace{-0.8cm} 
\includegraphics[height=8.5cm]{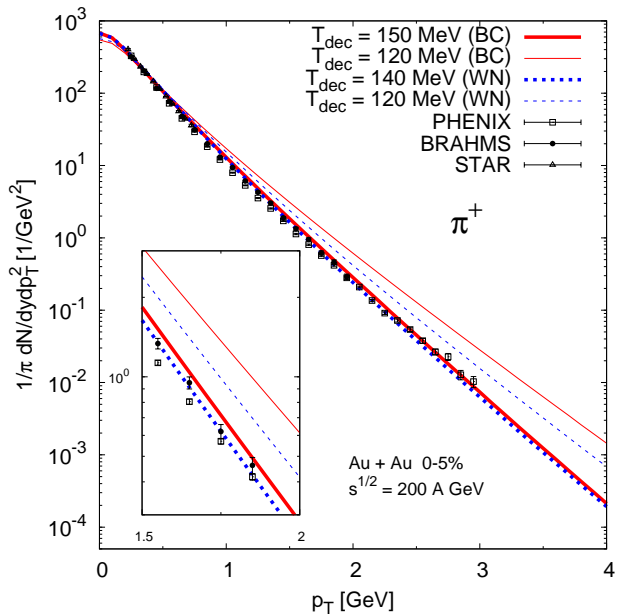} 
\label{spectraRHIC} 
\vspace{-0.4cm} 
\caption{\protect\small Transverse momentum spectra of positive pions at $y\sim 0$ in 5 \% most 
central RHIC 200 GeV Au+Au collisions, computed from constant-temperature decoupling boundaries for 
different initial states. The data shown are from 
BRAHMS~\cite{Bearden:2004yx}, PHENIX~\cite{Adler:2003cb} and STAR~\cite{Adams:2003xp} collaborations.}
\vspace{-0.4cm} 
\end{figure} 
%%%%%%%%%%%%%%%%%%%%% FIGURE %%%%%%%%%%%%%%%%%%%%%%%%%%%%%%%% 
%%%%%%%%%%%%%%%%%%%%% FIGURE %%%%%%%%%%%%%%%%%%%%%%%%%%%%%%%% 
\begin{figure}[t] 
%\vspace{-0.6cm} 
%\hspace{-0.8cm} 
\includegraphics[height=8.5cm]{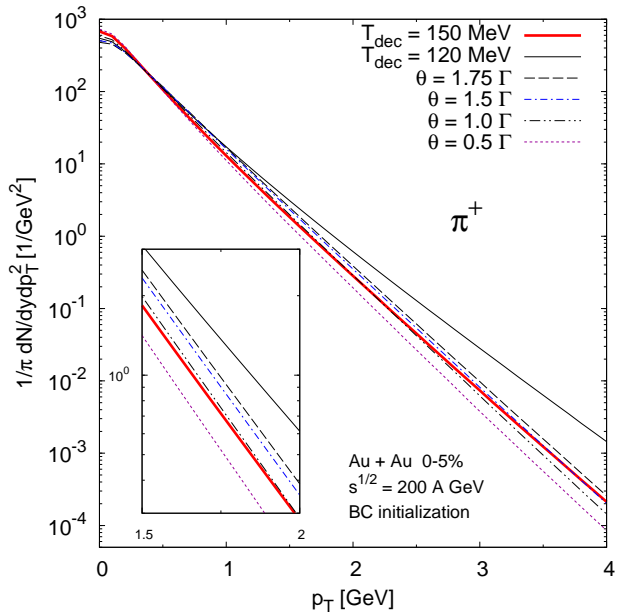} 
\label{spectraRHIC_BC} 
\vspace{-0.4cm} 
\caption{\protect\small Transverse momentum spectra of positive pions at $y\sim0$ in 5 \% most 
central RHIC 200 GeV Au+Au collisions from the dynamical decoupling condition for different 
values of the constant $c$, compared with the constant-temperature 
spectra. The calculation is made for the BC initial conditions.} 
\vspace{-0.4cm} 
\end{figure} 
%%%%%%%%%%%%%%%%%%%%% FIGURE %%%%%%%%%%%%%%%%%%%%%%%%%%%%%%%% 
%%%%%%%%%%%%%%%%%%%%% FIGURE %%%%%%%%%%%%%%%%%%%%%%%%%%%%%%%% 
\begin{figure}[bht] 
%\vspace{-0.6cm} 
%\hspace{-0.8cm} 
\includegraphics[height=8.5cm]{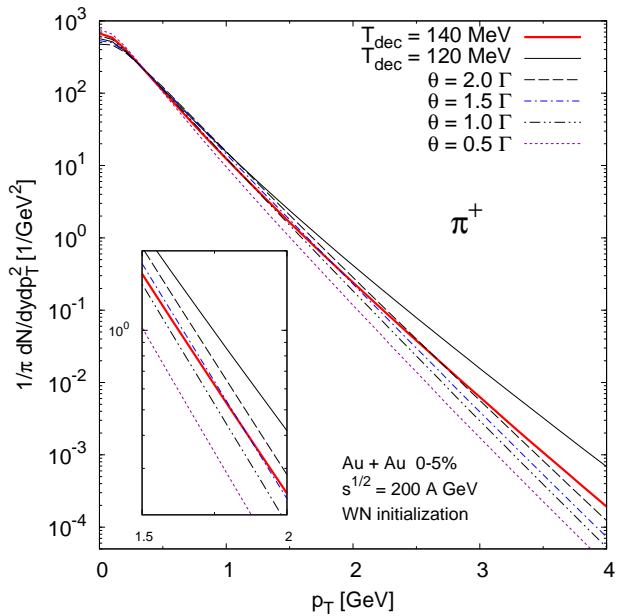} 
\label{spectraRHIC_WN} 
\vspace{-0.4cm} 
\caption{\protect\small As Fig.~5 but for WN initial conditions.} 
\vspace{-0.4cm} 
\end{figure} 
%%%%%%%%%%%%%%%%%%%%% FIGURE %%%%%%%%%%%%%%%%%%%%%%%%%%%%%%%% 

In Fig.~4 we show our previous results with BC initial profile~\cite{Eskola:2005ue} for 5 \% most central 
Au+Au collisions at $\sqrt{s_{NN}} = 200$ GeV,  compared with the RHIC data from BRAHMS~\cite{Bearden:2004yx}, 
PHENIX~\cite{Adler:2003cb} and STAR~\cite{Adams:2003xp} collaborations. 
With the WN profile for the initial energy and net-baryon number densities, the pressure gradients 
are smaller and the transverse flow builds up more slowly. Thus a lower $T_{\rm dec}$ is 
needed with the WN initial profile in order to reproduce the RHIC data.
The figure demonstrates that both the BC initial conditions with freeze-out at $T_{\rm dec} = 150$ MeV
and the WN initial conditions with freeze-out at $T_{\rm dec} = 140$ MeV are in good agreement with the 
data and with each other. We shall use these constant-$T_{\rm dec}$ results as a comparison baseline in 
what follows. Note how the freeze-out at $T_{\rm dec} = 120$ MeV clearly overshoots the data, for both initial conditions.

Figure~5 shows the $p_T$-spectra of positive pions, obtained with the BC initialization and using
the dynamical decoupling Eq. (\ref{dec_crit}). The BC-initialized $T_{\rm dec}=150$ and 120 MeV
results from Fig.~4 are shown for comparison. Figure 6 shows the corresponding results with the WN initialization.
We notice that for the BC initial condition the $c$ values in the range $1 - 1.5$
give a reasonable description of the data, the spectra being all close to our previous
calculation for $T_{\rm dec} = 150$ MeV. A closer inspection indicates that $c \simeq 1$ gives perhaps 
the best description. The fact that the constant-$T_{\rm dec}$ leads to very similar spectra as the rate 
condition with $c\sim 1$, which we consider to be physically better motivated, shows that also the 
constant-$T_{\rm dec}$ on the average correctly captures the effects of the rates competition. 
As indicated by Fig.~6, the use of WN profile supports slightly larger values of $c$, with 
$c \simeq 1 - 1.5$  giving the best description of the data at $p_T\lsim 2$ GeV, i.e. in the most 
relevant region for hydrodynamical models. On the basis of RHIC data we can thus conclude that within our 
dynamical freeze-out condition, a good description of the data is obtained with $c \simeq 1 - 1.5$. 

%%%%%%%%%%%%%%%%%%%%% FIGURE %%%%%%%%%%%%%%%%%%%%%%%%%%%%%%%% 
\begin{figure}[bht] 
\vspace{-0.0cm} 
\hspace{-0.0cm} 
\includegraphics[height=8.5cm]{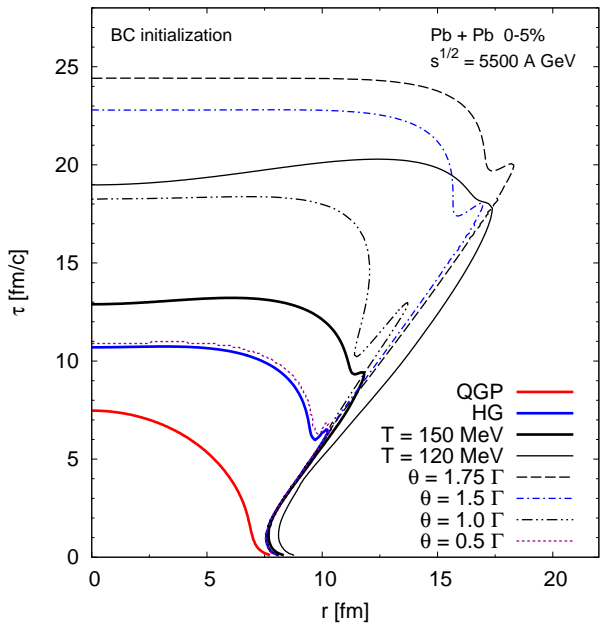} 
\label{flowLHC_BC} 
\vspace{-0.3cm} 
\caption{\protect\small  As Fig.~3 but for 5 \% most central LHC 5500 GeV Pb+Pb collision.} 
\vspace{-0.5cm} 
\end{figure} 
%%%%%%%%%%%%%%%%%%%%% FIGURE %%%%%%%%%%%%%%%%%%%%%%%%%%%%%%%% 
\section{LHC results}

In constructing the hypersurface for the Cooper-Frye decoupling based on the dynamical freeze-out 
condition, we assume that freeze-out dynamics is governed by local conditions only. Thus we expect 
that the constant $c$ obtained above is independent of the global properties of the system, such as 
the system size or the centrality of the collision. In particular, we expect $c$ to be independent of 
the collision energy. Therefore we can use the value of $c$, which is determined from the RHIC data, 
to predict the decoupling boundary at the LHC. 

Figure~7 shows the phase boundaries and decoupling curves for 5 \% most central Pb+Pb collisions at 
the LHC energy $\sqrt{s_{NN}}=5.5$~TeV, in the case of the BC profile for the initial energy density. The 
behaviour of the decoupling boundaries, which are obtained from the dynamical decoupling condition, 
is very similar to the RHIC results above: during the early times of the evolution decoupling near 
the spatial boundary of the system occurs immediately after hadronization is complete, while at the 
center of the fireball the system decouples at later times, when decoupling temperature is 
significantly lower. The decoupling boundaries for the WN profile are qualitatively similar to the BC case, 
and therefore not shown. 
%%%%%%%%%%%%%%%%%%%%% FIGURE %%%%%%%%%%%%%%%%%%%%%%%%%%%%%%%% 
\begin{figure}[bht] 
%\vspace{-0.6cm} 
%\hspace{-0.8cm} 
\includegraphics[height=8.5cm]{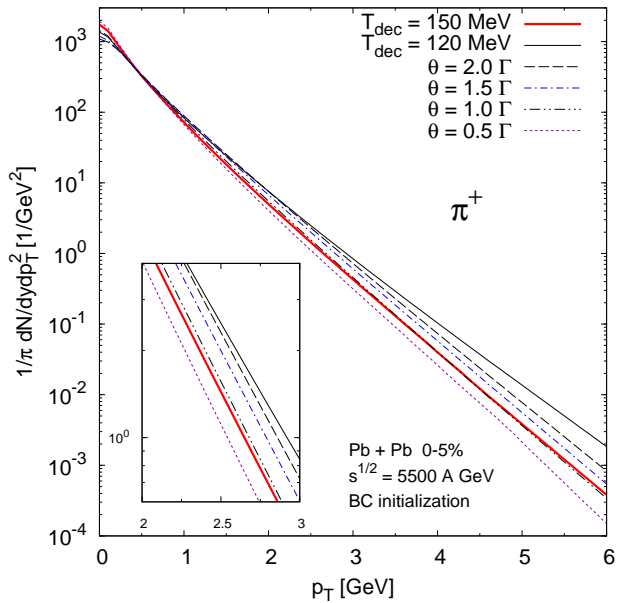} 
\label{spectraLHC_BC} 
\vspace{-0.4cm} 
\caption{\protect\small As Fig.~5 but for 5 \% most central LHC 5500 GeV Pb+Pb collision.} 
\vspace{-0.4cm} 
\end{figure} 
%%%%%%%%%%%%%%%%%%%%% FIGURE %%%%%%%%%%%%%%%%%%%%%%%%%%%%%%%% 
%%%%%%%%%%%%%%%%%%%%% FIGURE %%%%%%%%%%%%%%%%%%%%%%%%%%%%%%%% 
\begin{figure}[bht] 
%\vspace{-0.6cm} 
%\hspace{-0.8cm} 
\includegraphics[height=8.5cm]{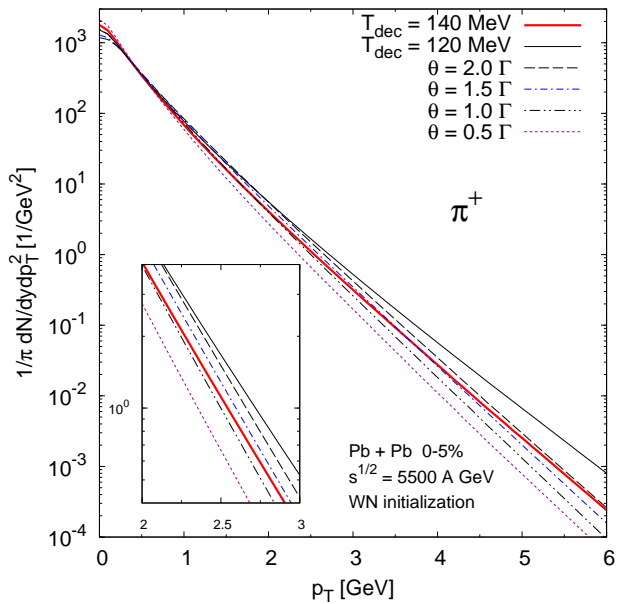} 
\label{spectraLHC_WN} 
\vspace{-0.4cm} 
\caption{\protect\small As Fig.~6 but for 5 \% most central LHC 5500 GeV Pb+Pb collision.} 
\vspace{-0.4cm} 
\end{figure} 
%%%%%%%%%%%%%%%%%%%%% FIGURE %%%%%%%%%%%%%%%%%%%%%%%%%%%%%%%% 

Transverse momentum spectra of positive pions in 5 \% most central Pb+Pb collisions at the LHC, 
calculated from the decoupling boundaries above are shown in Fig.~8 for the BC and Fig.~9 for the WN 
initializations. Fixing the proportionality constant $c$ between the expansion and scattering rates 
from the RHIC data in Fig.~4, our prediction for the positive pion $p_T$ spectra is given by the 
curves with $c=1-1.5$. Also the constant-$T_{\rm dec}$ results with $T_{\rm dec} = 150$ and 120 MeV are shown 
for the BC initialization in Fig.~8, and with $T_{\rm dec} = 140$ and 120 MeV for WN 
initialization in Fig.~9. We make the following interesting observations:

\emph{(i)} Once the value of $c$ is fixed from RHIC data, take e.g. $c=1$ with the BC initialization 
(Figs. 4 and 8), the dynamical decoupling results both at RHIC and LHC are reproduced with the same
$T_{\rm dec}=150$~MeV. This suggests that $T_{\rm dec}$ does not change considerably from RHIC to LHC.
The same holds for the results for WN initialization, with effective  $T_{\rm dec}=140$~MeV.
Upper limit $c = 1.5$ corresponds a slightly smaller effective $T_{\rm dec}$ in both cases. Thus, in our hydrodynamical
framework, we expect effective $T_{\rm dec}$ to be in the range $140-150$ MeV.

\emph{(ii)} The LHC pion spectra obtained with $c=1-1.5$ or with an effective $T_{\rm dec}=150$~MeV for
the BC case in Fig. 8, or with an effective $T_{\rm dec}=140$~MeV for the WN case in Fig. 9, are 
all very close to each other. This shows that our LHC predictions for pion spectra 
\cite{Eskola:2005ue} are not very sensitive to the uncertainties of the initial profile, provided that the 
RHIC data have been exploited to constrain the decoupling dynamics.

\emph{(iii)} As seen in Figs. 4-6, the RHIC data outrules the values of $c$ which would be much 
larger than 1-2. Correspondingly, as shown before \cite{Eskola:2002wx,Eskola:2005ue}, 
a low effective constant $T_{\rm dec}$, cannot be used at RHIC.
The present study of local dynamic decoupling condition lends support to the prediction that 
if constant-$T_{\rm dec}$ decoupling is used, low $T_{\rm dec}$ values are unlikely also at
the LHC.

\emph{(iv)} Intuitively, since the flow develops faster already in the plasma phase at the LHC than 
at RHIC, one might anticipate the decoupling at the LHC to happen at higher effective 
temperatures than at RHIC. The present study, especially the BC case, now shows that the effective 
decoupling temperature in central Pb+Pb collisions at the LHC is in fact quite close to that in 
central Au+Au collisions at RHIC.

\section{Conclusions}
We have computed the transverse momentum spectra of pions in nearly central Au+Au collisions
at RHIC and Pb+Pb collisions at the LHC within the framework of ideal 
hydrodynamics, by taking the initial conditions from a pQCD + saturation model.
In particular, we have applied a dynamical decoupling condition \cite{Dumitru:1999ud,Hung:1997du}, 
where we compare the local expansion rate to the local pion-pion scattering rate.
The proportionality constant between the two rates at decoupling is extracted from the RHIC data
on pion transverse momentum spectra. The same proportionality constant is then used to present  
predictions of pion spectra at the LHC. 

Although the model is simple and we consider pion-pion collisions and pion spectra only, the fact 
that the proportionality constant obtained is of the order of unity, gives us some 
confidence that we have captured an essential part of the physics in the breakdown of 
the hydrodynamic stage of matter into final free hadrons.
The dynamical decoupling condition does not necessarily indicate the end 
of all hadronic interactions, but for $\pi\pi$ interactions this is expected to be the case, at least
in the sense that interactions cannot maintain local thermal equilibrium. Nucleons 
most likely decouple somewhat later than pions because of the strong $\Delta$ resonance in 
$\pi N$ interactions~\cite{Teaney:2001av, Hung:1997du}. For pions, 
however, this is only a minor effect, due to the small number of nucleons compared to pions. 
A consistent modelling of these interesting details is beyond the scope of the present paper.

Studying both binary collision and wounded nucleon profiles for the initial energy densities, we have 
compared the results obtained with the dynamical decoupling criterion to those obtained with  
effective constant freeze-out temperatures. We find that the effective freeze-out temperatures for 
pions at the LHC will be very close to those at RHIC, $T_{\rm dec}=140-150$ MeV. With the dynamical 
decoupling criterion, we also find that the hydrodynamic prediction of the pion $p_T$ spectra at 
the LHC is fairly insensitive to the details of initial transverse profile, once the total entropy 
is fixed and once the decoupling dynamics is first constrained by the RHIC data.

We thank P. Huovinen for discussions and the Academy of Finland, Project 115262, for financial 
support.

\end{document}